\documentclass[12pt]{article}

\usepackage[margin=1in]{geometry}

\usepackage{setspace}
\usepackage{cite}
\usepackage{amsmath,amssymb,amsfonts}
\usepackage{algorithmic}
\usepackage{graphicx}
\usepackage{textcomp}
\usepackage{lineno}
\usepackage{titlesec}
\usepackage{authblk}
\usepackage{float}
\usepackage{placeins}
\titleformat{\section}{\large\bfseries}{\thesection}{0.6em}{}
\titleformat{\subsection}{\normalsize\bfseries}{\thesubsection}{0.6em}{}

\title{\textbf{ALD Oxidant as a Tuning Knob for Memory Window Expansion in Ferroelectric FETs for Vertical NAND Applications}}

\date{}
\begin{document}

%\linenumbers
\title{ALD Oxidant as A Tuning Knob for Memory Window Expansion in Ferroelectric FETs for Vertical NAND Applications}

\author{Ranie Jeyakumar, Prasanna Venkatesan, Lance Fernandes, Salma Soliman, Priyankka Ravikumar, Taeyoung Song, Chengyang Zhang,  Woohyun Hwang, Kwangyou Seo, Suhwan Lim, Wanki Kim, Daewon Ha, Shimeng Yu, Suman Datta, and Asif Khan \thanks{This work was supported by Samsung Electronics (IO250304-12193-01). Device fabrication was performed at the Georgia Tech Institute for Matter and Systems (IMS), a facility supported by the NSF NNCI program (ECCS-1542174). }
\thanks{Ranie Jeyakumar, Prasanna Venkatesan, Lance Fernandes, Salma Soliman, Priyankka Ravikumar, Taeyoung Song, Chengyang Zhang, Shimeng Yu are with  the School of ECE, Georgia Institute of Technology, Atlanta,
GA 30318 USA (raniesj@gatech.edu) }
\thanks{ Woohyun Hwang, Kwangyou Seo, Suhwan Lim, Wanki Kim, Daewon Ha are with the Semiconductor Research and
Development, Samsung Electronics Company Ltd., Hwaseong 445701,
South Korea.
}
\thanks{Suman Datta and Asif Khan are with the School of ECE and the School
of Material Science and Engineering, Georgia Institute of Technology,
Atlanta, GA 30318 USA}}
\maketitle
\begin{abstract}
Dielectric inserts are widely used to expand the memory window (MW) in ferroelectric FETs (FeFETs) for vertical NAND applications, with prior efforts focused primarily on material selection and stack positioning. Here, we demonstrate that the ALD oxidant used for the Al$_2$O$_3$ interlayer serves as a process-level tuning knob for MW engineering. H$_2$O-grown Al$_2$O$_3$ yields a significantly larger MW ($\sim$7-8~V) compared to O$_3$ ($\sim$4~V) for both gate-injection (12/3) and tunnel dielectric (8/3/8) configurations. While the tunnel dielectric (8/3/8) stack maintains robust retention up to $10^4$~s at 125$^\circ$C despite the larger MW, the gate-injection (12/3) configuration exhibits pronounced retention degradation for the H$_2$O case. The enhanced MW is attributed to higher interlayer leakage associated with H$_2$O-based ALD. These results establish oxidant choice as a key process parameter for co-optimizing MW and retention in ferroelectric NAND technologies.

\end{abstract}

\section{Introduction}
\begin{figure}[H]
\centerline{\includegraphics[width=0.7\columnwidth]{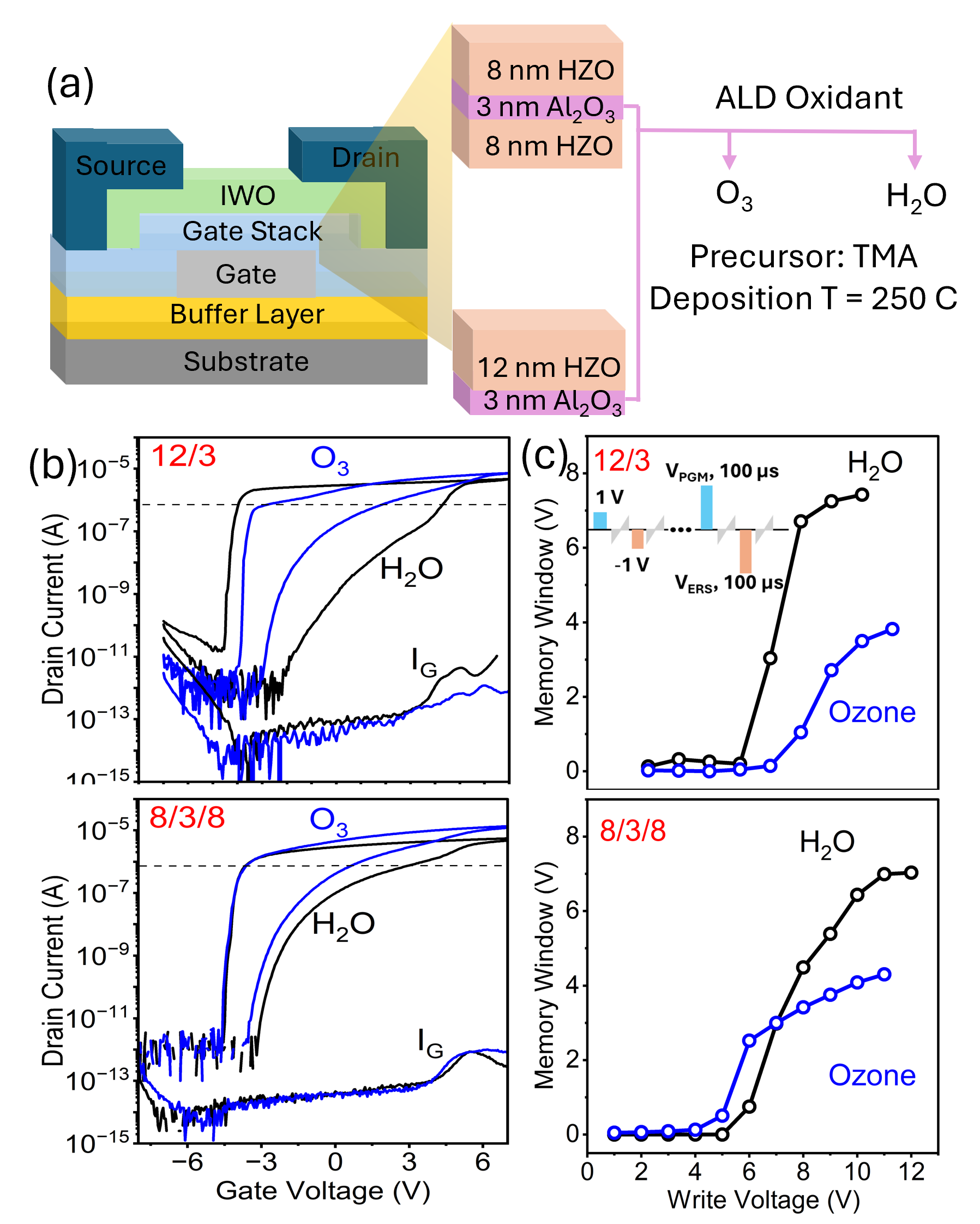}}
\caption{Schematic of the device structures with (a) 12/3 and (b) 8/3/8 gate stack is shown, (b) DC-IV characteristics of the two devices with 12/3 gate stack and 8/3/8 gate stack are shown. (c) MW as a function of V$_{WR}$ with pulse
width of 100 $\mu$s for all the stacks} 
\label{fig1}
\end{figure}
The exponential growth in data-intensive applications places unprecedented demands on ultra-high-density storage solutions. Charge-trap flash (CTF)--based 3D NAND has been central to meeting this demand, scaling beyond 200--300 layers and achieving data densities exceeding 20 GB/mm² through vertical integration and multi-level cell operation\cite{b0,b4}. However, as vertical scaling progresses, CTF NAND faces increasing challenges related to retention loss, lateral charge migration, rising program/erase voltages, and reliability degradation due to reduced vertical pitch \cite{b3}. These limitations motivate the exploration of alternative memory concepts that can extend vertical NAND scaling beyond current limits. Ferroelectric field-effect transistors (FeFETs) have emerged as promising candidates for next-generation non-volatile memory technologies offering a compelling pathway to overcome the fundamental limitations of conventional CTF-based memory as scaling continues \cite{b1,b2}.

For V-NAND applications, however, FeFETs must satisfy stringent
requirements, including a total gate-stack thickness below 20 nm and a large memory window (MW \textgreater{} 7.5 V) compatible with multi-bit operation at operating voltages below 15 V \cite{b5}. 
Dielectric inserts have emerged as an effective tuning knob for memory window (MW) expansion in ferroelectric NAND, with material composition, thickness scaling, and stack positioning widely explored \cite{b6,b7,lanceedl,lanceimw,b8}. However, for scalable manufacturing, insight into deposition process parameters is equally critical. In this work, we demonstrate that the ALD oxidant used for the dielectric insert serves as an additional process-level tuning knob for MW engineering, enabling performance modulation without altering the physical gate-stack geometry.

\section{Experimental Details}
%\label{sec:Experimental Details}

Two ferroelectric gate-stack configurations with a total thickness of 15 nm and 19 nm were implemented in 4\% W-doped In\textsubscript{2}O\textsubscript{3}-channel FeFETs: (1) a 3 nm Al\textsubscript{2}O\textsubscript{3} layer inserted between the gate metal and a 12 nm HZO layer (denoted as 12/3), and (2) a 3 nm Al\textsubscript{2}O\textsubscript{3} layer sandwiched between two 8 nm HZO layers (denoted as 8/3/8) as shown in Fig. 1(a). For both stack configurations, the Al\textsubscript{2}O\textsubscript{3} gate interlayer was deposited by thermal ALD using either H\textsubscript{2}O or O\textsubscript{3} as the oxidant. In all cases, the HZO layers were deposited by thermal ALD with O\textsubscript{3} as the oxidant. The detailed process flow is described in Ref. \cite{lanceimw}.
\section{Results and Discussions}
Fig. 1(b) shows the $I_D$–$V_{GS}$ characteristics of the devices measured at $V_{DS} = 50$ mV. Fig. 1(c) plots the memory window (MW) as a function of write voltage, extracted using a pulsed measurement setup with a pulse width of 100 $\mu$s. Both the 12/3 and 8/3/8 FeFETs incorporating H\textsubscript{2}O-grown Al\textsubscript{2}O\textsubscript{3} exhibit a maximum MW close to 8 V and 7 V, respectively, whereas devices with O\textsubscript{3}-grown Al\textsubscript{2}O\textsubscript{3} show a significantly smaller MW of approximately 4V.

\begin{figure}[H]
\centerline{\includegraphics[width=0.7\columnwidth]{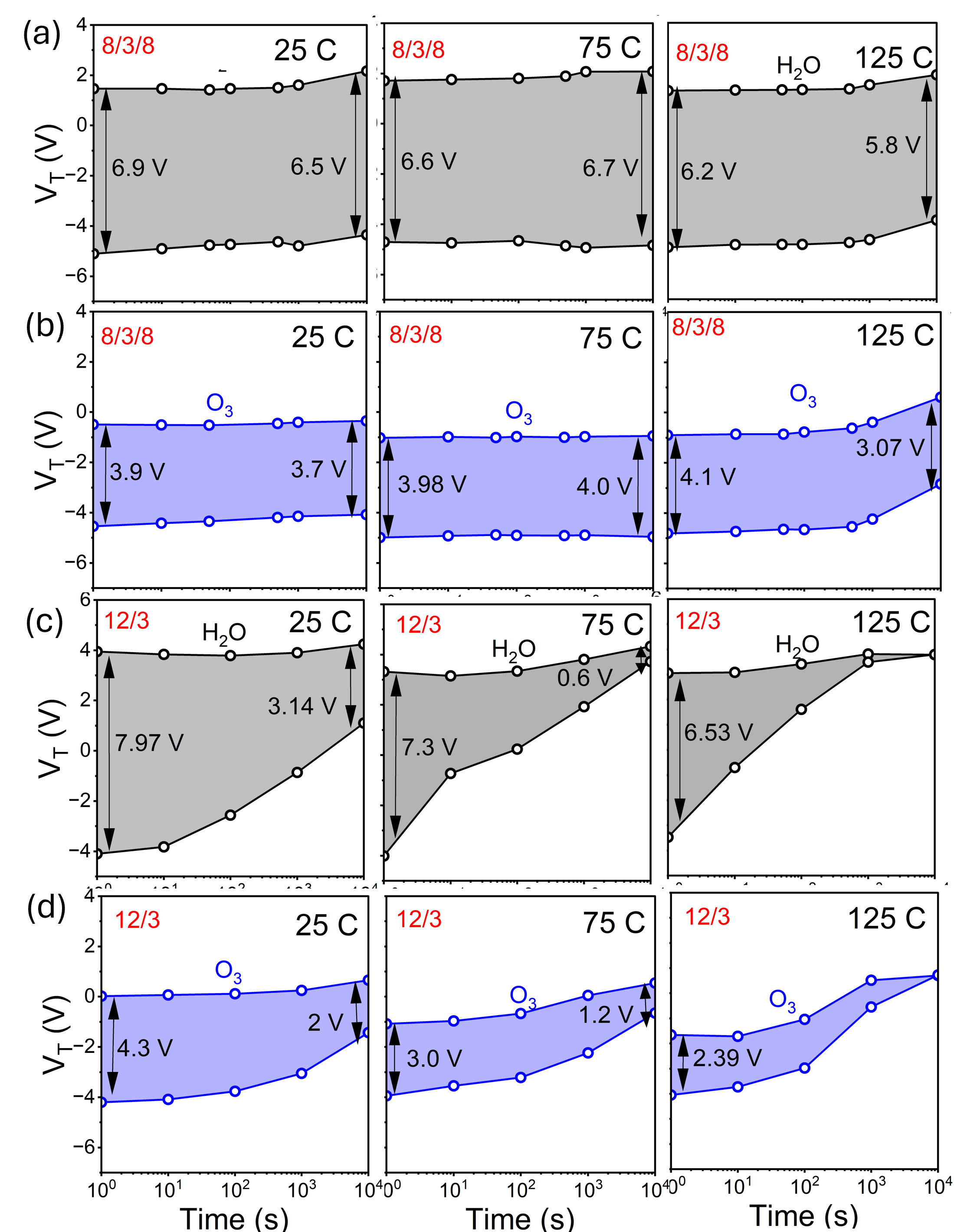}}
\caption{ Retention of FEFETs with the (a) 8/3/8 stack with H$_2$o, (b) 8/3/8 ozone, (c) 12/3  H$_2$o and (d) 12/3 ozone Al\textsubscript{2}O\textsubscript{3} layers are shown. Retention of the program and erase states is measured up to 10$^4$ s done at RT, 75 $^\circ$C, and 125 $^\circ$C.  }
\label{fig2}
\end{figure}
Figures 2(a–d) show the retention characteristics of all four FeFETs measured up to 10\textsuperscript{4} s at 25, 75, and 125~$^\circ$C. For the 8/3/8 configuration, the H$_2$O-grown device exhibits a reduction in MW from 6.2~V to 5.8~V at 125~$^\circ$C over time. Similarly, the ozone-grown device shows a decrease in MW from 4.1~V to 3.07~V at 125~$^\circ$C. The overall $V_T$ drift is comparable for both oxidants, indicating that retention in the 8/3/8 stack is largely insensitive to the interlayer deposition chemistry.

In contrast, the 12/3 stack (Fig.~2(c–d)) shows strong oxidant dependence. The H$_2$O-grown Al$_2$O$_3$ device exhibits pronounced degradation, with the MW decreasing from 7.97~V to 3.14~V at 25~$^\circ$C. The ozone-grown device starts with a smaller MW (4.3~V at 25~$^\circ$C) and reduces to 2~V, with comparatively lower $V_T$ drift. Overall, while both Al$_2$O$_3$ interlayers yield similar retention behavior in the 8/3/8 configuration, the 12/3 stack shows significantly higher retention loss for the H$_2$O-grown interlayer compared to ozone. The retention behavior observed in these four gate stacks is consistent with prior reports \cite{b8}. In the 12/3 FeFETs, charges accumulated at the HZO/Al\textsubscript{2}O\textsubscript{3} interface can tunnel through the Al\textsubscript{2}O\textsubscript{3} layer, leading to a reduction in the memory window over time. In contrast, in the 8/3/8 FeFETs, the thicker HZO layer effectively confines these charges, resulting in improved charge retention and a more stable memory window.
%Fig. 2(e) shows a benchmark plot of retention characteristics. 

\begin{figure}[H] \centerline{\includegraphics[width=\columnwidth]{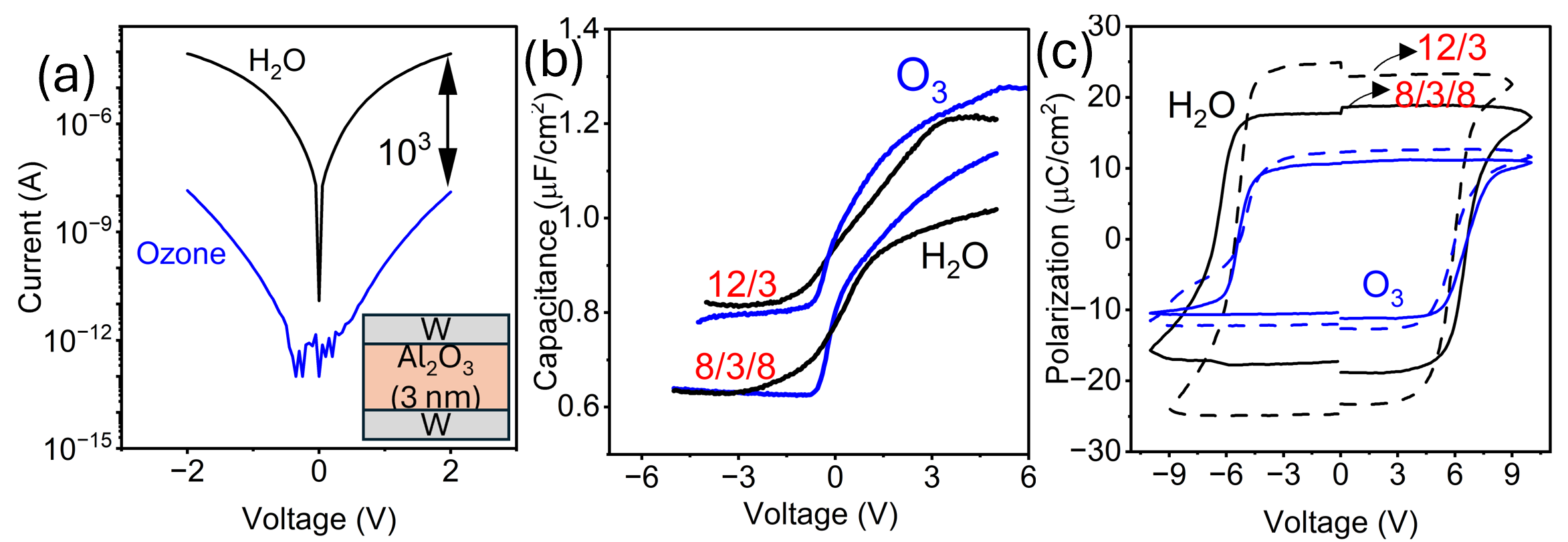}} \caption{ (a) Leakage characteristics of the Al\textsubscript{2}O\textsubscript{3} layer measured from
MIM deposited through H$_2$O and ozone, C-V characteristics of the (b) 12/3 and 8/3/8 devices with both layers showing similar dielectric constant, PUND
measurements in (c) 12/3 and 8/3/8 FeFETs showing higher 2P$_r$ for the device with H$_2$O Al\textsubscript{2}O\textsubscript{3} layer } \label{fig2} \end{figure}

To elucidate the origin of the MW dependence on the Al\textsubscript{2}O\textsubscript{3} deposition technique—independent of its position within the gate stack—we further investigate leakage, capacitance, and switched polarization in these stacks. Standalone W--Al$_2$O$_3$ (3~nm)--W capacitors were fabricated using thermal ALD with either H$_2$O or O$_3$ as the oxidant, and the corresponding current-voltage characteristics are shown in Fig. 3(a). Gate capacitance ($C_G$)--gate voltage ($V_G$) characteristics and polarization--voltage ($P$--$V_G$) characteristics of the FeFETs were measured, as shown in Fig.~3(b) and Fig.~3(c), respectively.

As shown in Fig.~3(a), the H$_2$O-ALD Al$_2$O$_3$ sample exhibits orders-of-magnitude higher leakage current compared to the O$_3$-ALD sample. This observation is consistent with prior reports indicating that ozone-based ALD reduces hydrogen and hydroxyl incorporation, resulting in significantly lower leakage currents \cite{b12,b13}. In contrast, Fig.~3(b) shows no significant dependence of the gate capacitance ($C_G$) on the Al$_2$O$_3$ oxidant for either the 8/3/8 or 12/3 configurations. The extracted dielectric constants of Al$_2$O$_3$ are 8.4 and 9 for H$_2$O- and O$_3$-based processes, respectively. 

The memory window can be expressed as $\mathrm{MW} = \frac{P_r - Q_{it}}{C_{\mathrm{FE}}} + \frac{Q_{it'} - Q_{it}}{C_{\mathrm{IL}}}$, where $P_r$ is the remanent polarization of the ferroelectric layer, $Q_{it}$ is the screening charge density at the FE--channel interface, and $Q_{it'}$ is the trapped charge density at the FE/IL interface that contributes to MW enhancement. Given the small difference in dielectric constant (8.4 vs.~9), variations in permittivity of Al$_2$O$_3$  alone cannot account for the substantial MW difference observed between the two oxidants.
\begin{figure}[H] \centerline{\includegraphics[width=\columnwidth]{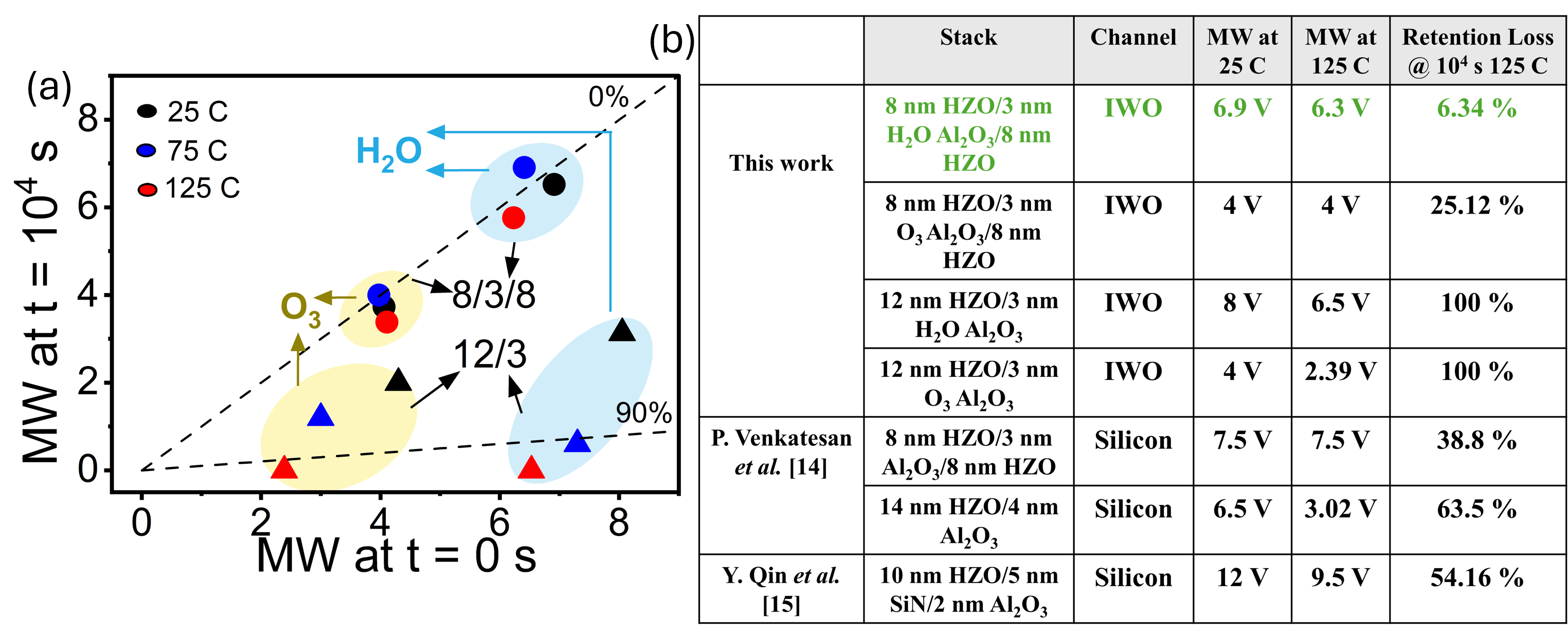}} \caption{ (a) Memory window (MW) at $t = 10^{4}$ s versus initial MW ($t = 0$ s) for 8/3/8 (circles) and 12/3 (triangles) FeFET stacks with H$_2$O and O$_3$ Al$_2$O$_3$ interlayers measured at 25~$^\circ$C, 75~$^\circ$C, and 125~$^\circ$C. The dashed line indicates ideal retention (0\% loss). (b) Benchmark comparison of MW and retention loss with prior FeFET reports. } \label{fig2} \end{figure}
Based on these observations, we hypothesize that the higher leakage associated with the H$_2$O-ALD interlayer facilitates the observed enhancement in MW. In the 12/3 structure, the increased leakage promotes stronger gate-side injection, enabling a larger amount of charge to accumulate and subsequently convert into trapped charge ($Q_{it'}$) at the FE/IL interface. This additional trapped charge contributes directly to MW expansion. However, the same elevated leakage also accelerates charge loss, resulting in degraded retention for the H$_2$O-grown Al$_2$O$_3$ compared to its O$_3$ counterpart in the 12/3 configuration.

In the 8/3/8 structure, where the Al$_2$O$_3$ interlayer is sandwiched between two HZO layers, higher leakage facilitates improved polarization switching, leading to a larger remanent polarization ($P_r$), which further contributes to the increased MW. The retention remains largely comparable for both interlayers in the 8/3/8 device as the charge leakage from the HZO/Al\textsubscript{2}O\textsubscript{3} interface is primarily governed by the HZO layers rather than the interlayer itself \cite{b14}. As a result, variations in Al$_2$O$_3$ leakage have a minimal impact on the overall retention characteristics, leading to similar retention behavior for both oxidants, as observed in Fig. 4.

\section{Conclusion}
In summary, we demonstrate that the choice of ALD oxidant for the Al$_2$O$_3$ dielectric insert strongly influences memory window (MW) expansion in ferroelectric FETs for vertical NAND applications. H$_2$O-based ALD consistently yields a significantly larger MW compared to O$_3$ for both the 12/3 (gate-injection) and 8/3/8 (tunnel dielectric) configurations. While the 8/3/8 stack shows no substantial retention degradation up to $10^4$~s at 125~$^\circ$C despite the larger MW, the 12/3 configuration exhibits pronounced retention loss for the H$_2$O-grown interlayer. We hypothesize that the enhanced MW originates from the higher leakage associated with H$_2$O-grown Al$_2$O$_3$, which promotes increased polarization switching. These results establish oxidant choice as a process-level tuning knob for MW engineering, highlighting the importance of deposition chemistry in co-optimizing performance and retention in FE-NAND technologies.

\end{document}